\documentclass[twocolumn]{arxiv}

\title{Design And Optimization Of Multi-rendezvous Manoeuvres Based On Reinforcement Learning And Convex Optimization}

\usepackage[
    backend=biber,
    style=ieee,
    citestyle=numeric-comp
]{biblatex}
\newif\ifuniqueAffiliation
\uniqueAffiliationtrue

\ifuniqueAffiliation 
\author{%
    \href{https://orcid.org/0009-0006-5963-9927}{\includegraphics[scale=0.06]{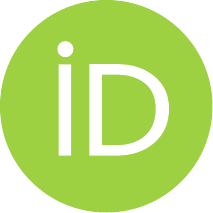}\hspace{1mm}Antonio López Rivera}\thanks{Permanent email: \texttt{antonlopezr99@gmail.com}.} \\
    MSc Spaceflight Dynamics \\
	Delft University of Technology\\
	Delft, The Netherlands \\
	\texttt{a.lopezrivera@student.tudelft.nl} \\
	\And
	Lucrezia Marcovaldi \\
	Department of AOCS-GNC\\
	Sener Aerospace \& Defence\\
	Tres Cantos, Madrid, Spain \\
	\texttt{lucrezia.marcovaldi@aero.sener} \\
        \And
	Jesús Ramírez \\
	Department of AOCS-GNC\\
	Sener Aerospace \& Defence\\
	Tres Cantos, Madrid, Spain \\
	\texttt{jesus.ramirez@aero.sener} \\
        \And
	Alex Cuenca \\
	Department of AOCS-GNC\\
	Sener Aerospace \& Defence\\
	Tres Cantos, Madrid, Spain \\
	\texttt{alex.cuenca@aero.sener} \\
        \And
	David Bermejo \\
	Department of AOCS-GNC\\
	Sener Aerospace \& Defence\\
	Tres Cantos, Madrid, Spain \\
	\texttt{david.bermejo@aero.sener} \\
}
\else
\usepackage{authblk}

\newbox{\orcid}\sbox{\orcid}{\includegraphics[scale=0.06]{orcid.pdf}} 
\author[1,2]{%
	\href{https://orcid.org/0009-0006-5963-9927}{\usebox{\orcid}\hspace{1mm}Antonio López Rivera\thanks{Permanent email: \texttt{antonlopezr99@gmail.com}}}%
}
\author[2]{%
	Lucrezia Marcovaldi\thanks{\texttt{lucrezia.marcovaldi@aero.sener}}%
}
\author[2]{%
	Jesús Ramírez\thanks{\texttt{jesus.ramirez@aero.sener}}%
}
\author[2]{%
	Alex Cuenca\thanks{\texttt{alex.cuenca@aero.sener}}%
}
\author[2]{%
	David Bermejo\thanks{\texttt{david.bermejo@aero.sener}}%
}
\affil[1]{Delft University of Technology, Delft, The Netherlands}
\affil[2]{Department of AOCS-GNC, Sener Aerospace \& Defence, Tres Cantos, Madrid, Spain}
\fi

\renewcommand{\headeright}{Presented at IAC 2024}
\renewcommand{\undertitle}{Manuscript presented at the 75\textsuperscript{th} International Astronautical Congress (IAC)\par Milan, Italy, 14-18 October 2024}
\renewcommand{\shorttitle}{RL and SCP for MRTO}

\hypersetup{
pdftitle={Design And Optimization Of Multi-rendezvous Manoeuvres Based On Reinforcement Learning And Convex Optimization},
pdfsubject={q-bio.NC, q-bio.QM},
pdfauthor={López Rivera et al.},
pdfkeywords={Trajectory Optimization, On-Orbit Servicing, Reinforcement Learning, Sequential Convex Programming},
}

\begin{document}

\twocolumn[

\maketitle

\begin{abstract}
Optimizing space vehicle routing is crucial for critical applications such as on-orbit servicing, constellation deployment, and space debris de-orbiting. Multi-target Rendezvous presents a significant challenge in this domain. This problem involves determining the optimal sequence in which to visit a set of targets, and the corresponding optimal trajectories: this results in a demanding NP-hard problem.
We introduce a framework for the design and refinement of multi-rendezvous trajectories based on heuristic combinatorial optimization and Sequential Convex Programming. Our framework is both highly modular and capable of leveraging candidate solutions obtained with advanced approaches and handcrafted heuristics. 
We demonstrate this flexibility by integrating an Attention-based routing policy trained with Reinforcement Learning to improve the performance of the combinatorial optimization process. 
We show that Reinforcement Learning approaches for combinatorial optimization can be effectively applied to spacecraft routing problems. 
We apply the proposed framework to the UARX Space OSSIE mission: we are able to thoroughly explore the mission design space, finding optimal tours and trajectories for a wide variety of mission scenarios. 
\end{abstract}
]

\saythanks{}


\section*{Nomenclature}

\subsection*{Symbols}


\tablefirsthead{%
}
\begin{supertabular}{ll}
$\omega$     & Argument of Perigee \\
$\mu$        & Earth gravitational constant \\
$R_e$        & Earth mean equatorial radius \\
$e$          & Eccentricity \\
$i$          & Inclination \\
$n$          & Orbital mean motion \\
$\Omega$     & Right Ascension of Ascending Node \\
$J_2$        & Second Earth zonal harmonic \\
$p$          & Semi-latus rectum \\
$a$          & Semi-major axis \\
$I_{sp}$     & Specific impulse \\
$g_0$        & Standard Earth gravity \\
$\sigma$     & True Anomaly \\
$L$          & True Longitude \\
\end{supertabular}

\subsection*{Acronyms}

\tablefirsthead{%
}
\begin{supertabular}{ll}
Advantage Actor-Critic                     & A2C \\
Argument Of Perigee                        & AOP \\
Deep Neural Network                        & DNN \\
Graph Attention Network                    & GAT \\
Machine Learning                           & ML \\
Mixed-Integer Nonlinear Programming        & MINLP \\
Modified Equinoctial Elements              & MEEs \\
Multiple Hohmann Transfer                  & MHT \\
Neural Combinatorial Optimization          & NCO \\
Nodal Inclination Change manoeuvre         & NIC \\ 
Operations Research                        & OR \\
Orbit Transfer Vehicle                     & OTV \\
Proximal Policy Optimization               & PPO \\
Reinforcement Learning                     & RL \\
Right Ascension of Ascending Node          & RAAN \\
SENER Optimization Toolbox                 & SOTB \\
Semi-major Axis                            & SMA \\
Sequential Convex Programming              & SCP \\
Space Traveling Salesman Problem           & STSP \\
Time Of Flight                             & TOF \\
Traveling Salesman Problem                 & TSP \\
Vehicle Routing Problem                    & VRP \\
\end{supertabular}

\section{Introduction}

The present work introduces a general optimization framework for multiple-rendezvous manoeuvres, which see a spacecraft approaching a sequence of objects in orbit as efficiently as possible. An optimal solution to the multi-target rendezvous trajectory optimization problem or STSP consists of the optimal sequence in which to visit a set of targets and the optimal transfer trajectory between each target in the optimal sequence. The STSP is an NP-hard MINLP problem with factorial complexity over the number of targets. The STSP bears resemblance to the classical TSP, albeit with the added complexities inherent to the space environment, notably a 6-dimensional state space, mass dynamics, propulsion constraints, and the change over time of target states due to secular perturbations, chiefly $J_2$ for Earth-orbiting spacecraft.

Formally, the STSP is the problem of finding a minimum weight path (if the spacecraft must end the tour back at its initial state, a Hamiltonian path) in a complete weighted graph $G\coloneqq \{\mathcal{V}(t), \mathbf{W}(\pi)\}$, where $\mathcal{V}(t)$ is the set of graph vertexes (targets, the state of which drifts over time) and $\mathbf{W}(\pi)\coloneqq\mathcal{V}\times\mathcal{V}\rightarrow\mathbb{R}^+$ is a map that associates an edge weight (a transfer cost) to each ordered vertex pair \cite{izzo_evolving_2015}, and may depend on the sequence $\pi$ in which the targets are visited. One such case is when payload mass is a large percentage of the spacecraft's wet mass, and thus deployment sequence has a non-negligible impact on fuel consumption.
The STSP is an example of a MINLP problem, which are notoriously difficult to approach, and has seen a considerable surge in interest in active debris removal missions \cite{izzo_evolving_2015,ricciardi_solving_2019,federici_evolutionary_2021,medioni_trajectory_2023,barea_large-scale_2020,narayanaswamy_low-thrust_2023} to tackle the space debris problem \cite{mark_review_2019,bonnal_active_2013}, as well as on-orbit servicing missions \cite{sellmaier_-orbit_2010,sarton_du_jonchay_-orbit_2022,federici_evolutionary_2021} and advanced space logistics concepts \cite{sorenson_multi-orbit_2023}. A standard approach to solve MINLP problems is Benders decomposition \cite{barea_large-scale_2020,morante_survey_2021}, where the MINLP problem is divided into a higher-level combinatorial optimization problem and a lower-level trajectory optimization problem; the higher level combinatorial problem is then solved exactly by applying Benders optimality cuts. Decomposition approaches using heuristic optimization are common as well \cite{federici_evolutionary_2021,medioni_trajectory_2023,narayanaswamy_low-thrust_2023}. In both approaches a transfer cost estimator is used to calculate the cumulative cost of tours in the combinatorial optimization problem. Transfer cost estimators may be database-dependent \cite{petropoulos_gtoc9_2017,lu_gtoc_2023}, database-independent (analytical), or learning-based \cite{li_deep_2020}. 

State-of-the-art combinatorial optimization methods fall in two camps: exact methods and heuristic methods, which are less costly and can produce near-optimal results, but cannot offer optimality guarantees whatsoever \cite{izzo_evolving_2015,francois_how_2019}. Exact methods based on tree searches are the norm for highly complex, large STSP variants; all winning submissions of the Global Trajectory Optimization Competitions have made use of tree search approaches \cite{izzo_evolving_2015,petropoulos_gtoc9_2017,hallmann_gtoc_2017}. 
Heuristic optimization methods however are an attractive option to solve smaller STSP instances (up to hundreds of targets \cite{izzo_evolving_2015}) due to their capacity to achieve near-optimal results with lower computational cost \cite{izzo_evolving_2015}, and are widely applied in the literature to tackle multi-rendezvous mission design \cite{izzo_evolving_2015,narayanaswamy_low-thrust_2023,medioni_trajectory_2023,federici_evolutionary_2021,ricciardi_solving_2019}.
Heuristic optimization methods have also been successfully applied to complex STSP instances where the cost of exact approaches is infeasible \cite{petropoulos_gtoc9_2017}.
Furthermore, the availability of highly performant, open-source heuristic multi-objective optimization libraries such as \texttt{pygmo}\footnote{\url{https://esa.github.io/pygmo2/}} \cite{biscani_parallel_2020} and \texttt{pymoo}\footnote{\url{https://pymoo.org/}} \cite{blank_pymoo_2020} greatly eases the application, benchmarking and selection of diverse heuristic optimization algorithms for specific problem variants.
We thus opt for population-based heuristic optimization to develop the combinatorial optimization component of our solver.

Convex optimization is leveraged to tailor optimal manoeuvres to the specific vehicle requirements, such as the specifics of the propulsion system. The applications of convex optimization in the field of aerospace have grown in importance in recent years. SOCP, in particular, is a common choice when nonlinear constraints are involved in the formulation of the OCP at the base of the orbit transfer optimization \cite{liu,domahidi}. These methods however are vulnerable to convergence issues: this has been tackled in the literature by using SCP solvers provided with an initial guess close to the optimal solution, and by keeping propagation and accumulation errors due to the iterations of the algorithm low; Foust et al. \cite{foust} and Ramírez and Hewing \cite{ramirez} propose the implementation of an SCP solver in combination with an \ordinalnum{4}-order Runge-Kutta, to integrate in the OCP an accurate model of the dynamics and reduce error accumulation. 
Discretization is crucial in determining the ability of the solver to find an optimal solution, re-adapting the dynamics and constraints of the OCP as functions of a finite number of parameters. 
Topputo et al. \cite{topputo1} develop an SCP solver based on mesh refinement to obtain a quasi-optimal warm start, demonstrating the robustness of the algorithm and its efficiency to generate warm starts for the SCP solver. SCP solvers based on interior-point methods, in particular, allow significant reduction in computational cost, making convex programming algorithms suitable also for demanding applications, such as on-line guidance \cite{ramirez,nocedal}.

This work presents a modular and adaptable STSP optimization framework based on Benders decomposition and heuristic combinatorial optimization, capable of solving highly tailored versions of the STSP. We demonstrate its capabilities by solving the multi-rendezvous trajectory optimization problem of UARX Space's OSSIE\footnote{\url{https://www.uarx.com/projects/ossie.php}} OTV: a translational and mass-dynamic multi-satellite deployment problem in LEO.

The paper is organized as follows: 
\autoref{sec: mission} describes the operational profile and performance envelope of the OSSIE OTV.
\autoref{sec:EnvironmentModel} defines the models used for the orbital dynamics and perturbations.
\autoref{sec: transfer} proceeds to define the trajectory design and trajectory cost estimation methods used for the UARX Space OSSIE OTV.
\autoref{sec: transfer} presents the Bolza formulation of the MINLP, the architecture of our optimization framework, introduces the heuristic combinatorial optimization methodology, and defines the SCP algorithm together with the OCP and the solver tailoring based on SOTB.
Experimental results are discussed in \autoref{sec:results}. \autoref{subsec:mission modelling} defines a model to generate randomized mission scenarios for OSSIE. \autoref{subsec: combinatorial results} and \autoref{subsec:rl results} discuss the results obtained from heuristic and neural combinatorial optimization. A statistical mission feasibility analysis follows in \autoref{subsec:mission analysis}. \autoref{subsec: results scp} presents the trajectory optimization results from SCP. The results section concludes with the formal verification of the obtained trajectories in a high-fidelity simulator in \autoref{subsec: verification results}.
Lastly, \autoref{sec:conclusion} lays out our main conclusions and recommendations for future research.

\section{Mission Profile}
\label{sec: mission}

\begin{figure}
    \centering
    \includegraphics[width=\linewidth]{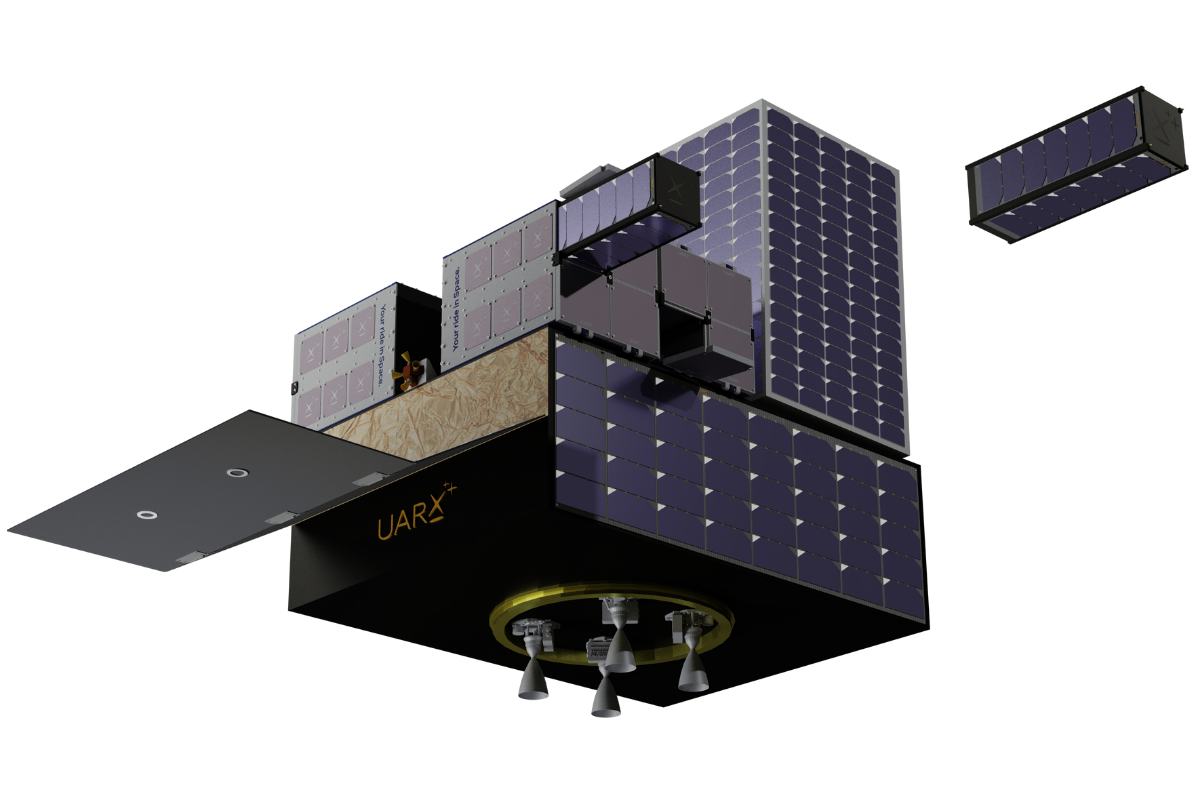}
    \caption{UARX Space OSSIE OTV. Credit: UARX Space.}
    \label{fig:OSSIE render}
\end{figure}

To demonstrate the capabilities of the proposed framework, the UARX Space Orbit Solutions to Simplify Injection and Exploration OTV, known as OSSIE, will be used as a case study. Depicted in \autoref{fig:OSSIE render}, OSSIE is a modular payload delivery platform with LEO, MEO and cis-lunar capability. In this paper, we consider a nominal mission profile aiming to deliver 4 PocketQubes, 8 CubeSats, and 1 small satellite to LEO. 

The propulsion system used for this mission consists of 4 parallel Dawn Aerospace B20 bi-propellant (nitrous oxide and propene) thrusters\footnote{\url{https://www.dawnaerospace.com/green-propulsion}} (specifications in \autoref{tab:OSSIE specs}).
As of the time of writing, the duty-cycle constraints of the thruster cluster constrain manoeuvre design to multiple-revolution transfers, with up to two impulses per orbit in LEO. 

\begin{table}
\caption{Specifications of the OSSIE OTV and Dawn Aerospace B20 thrusters.}
\label{tab:OSSIE specs}
\centering
\begin{adjustbox}{max width=\linewidth}
\begin{tabular}{llll}
\thickhline
OSSIE                     & Value & Dawn Aerospace B20            & Value \\ 
\hline
 Wet mass                 & \SI{235}{\kilo\gram} &Specific impulse           & \SI{277}{\second}   \\
 Payload mass             & \SI{80}{\kilo\gram} & Peak thrust                & \SI{12.6}{\newton}\\
 Fuel mass                & \SI{35}{\kilo\gram} & Minimum impulse bit        & \SI{1}{\newton\second}  \\
Cargo capacity           & 48 U  \\
\thickhline
\end{tabular}
\end{adjustbox}
\end{table}

The specifications of the current configuration of OSSIE follow in \autoref{tab:OSSIE specs}. OSSIE has a wet mass of approximately 235 kg, of which close to 50\% is shed through the mission ---either deployed or consumed propellant; this means that the mass deployment sequence may have a considerable impact on the fuel required to complete the mission. 
OSSIE is deployed to a nominal insertion orbit in LEO and  must conduct a decommissioning manoeuvre to ensure it decays within 5 years of EOL as per the ESA Space Debris Mitigation
Requirements \cite{esa_esa_2023}, meaning the start and end point of the transfer sequence are constrained.

Under the previous considerations, the OSSIE trajectory optimization problem results in a highly tailored multi-rendezvous trajectory optimization problem, which aims to minimize fuel consumption while accounting for:

\begin{itemize}
    \item Multi-revolution impulsive manoeuvres in LEO under perturbations.
    \item The impact of mass deployment sequence on propellant consumption.
    \item Insertion and decommissioning orbit constraints.
\end{itemize}

\section{Environment Model}
\label{sec:EnvironmentModel}

\begin{figure*}
    \begin{minipage}[t]{0.45\linewidth}
        \begin{figure}[H]
            \centering
            \includegraphics[width=1\linewidth]{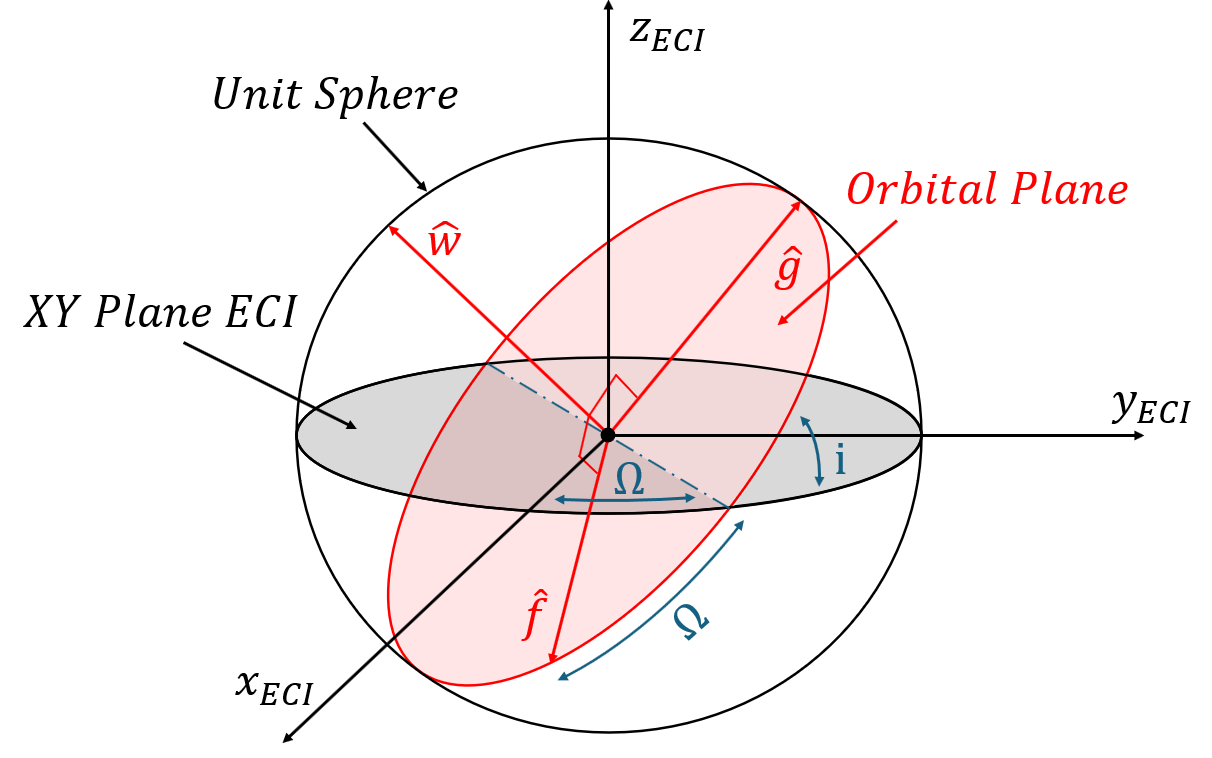}
            \caption{Modified Equinoctial Elements (MEE) with respect to orbital plane.}
            \label{fig:MEEs}
        \end{figure}
    \end{minipage}%
    \hspace{0.05\linewidth}%
    \begin{minipage}[t]{0.45\linewidth}
        \begin{figure}[H]
            \centering
            \includegraphics[width=1\linewidth]{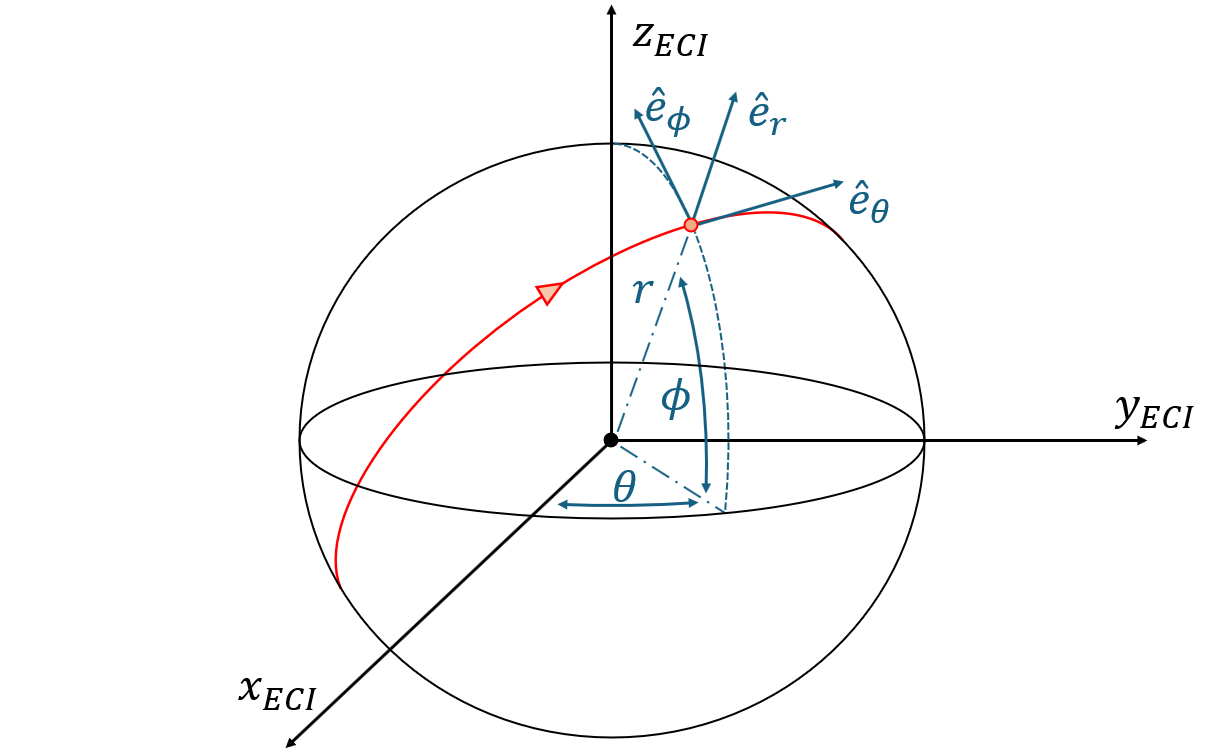}
            \caption{ECI and LVLH reference frames.}
            \label{fig:rsw reference frame}
        \end{figure}
    \end{minipage}
\end{figure*}

To set up the optimization space, the spacecraft state is modelled using the MEEs described by Hintz \cite{hintz_survey_2008}, including the retrograde factor $I$, which are nonsingular for all eccentricities and inclinations. The conversion from classical elements to MEEs follows in \autoref{eq:classical to mee}:

\begin{equation}
\label{eq:classical to mee}
\begin{aligned}
  & p = a(1-e^2) ;\\
  & f = e \cos(\omega + \Omega)   ;\\
  & g = e  \sin(\omega + \Omega)    ;\\
  & h = \tan(i/2)\sin(\Omega)    ;\\
  & k = \tan(i/2)\cos(\Omega)    ;\\
  & L = \theta + I\Omega + \omega    ;\\
\end{aligned}
\end{equation}

The Gauss Variational Equations for MEEs \cite{hintz_survey_2008} in \autoref{eq:EOM MEE}, are used to model the time evolution of the spacecraft's translational state:

\begin{equation}
\label{eq:EOM MEE}
    \begin{aligned}
    \frac{\mathrm{d}p}{\mathrm{d}t}=&\frac{2p}{w}\sqrt{\frac{p}{\mu}}\Delta_t;\\
    \frac{\mathrm{d}f}{\mathrm{d}t}=&\sqrt{\frac{p}{\mu}}\bigg\{\Delta_r\sin(L);\\
        &\left.+\frac{(w+1)\cos(L)+f}{w}\Delta_t-g\frac{v}{w}\Delta_n\right\};\\
    \frac{\mathrm{d}g}{\mathrm{d}t}=&\sqrt{\frac{p}{\mu}}\bigg\{-\Delta_r\cos(L);\\
        &\left.+\frac{(w+1)\sin(L)+g}{w}\Delta_t-f\frac{v}{w}\Delta_n\right\};\\
    \frac{\mathrm{d}h}{\mathrm{d}t}=&\sqrt{\frac{p}{\mu}}\frac{s^2}{2w}\cos(L)\Delta_n;\\
    \frac{\mathrm{d}k}{\mathrm{d}t}=&\sqrt{\frac{p}{\mu}}\frac{s^2}{2w}\sin(L)\Delta_n;\\
    \frac{\mathrm{d}L}{\mathrm{d}t}=&\sqrt{\mu p}\left(\frac{w}{p}\right)^2+ \sqrt{\frac{p}{\mu}}\frac{v}{w}\Delta_n;
    \end{aligned}
\end{equation}

\noindent where $s^2$, $v$ and $w$ are defined as follows,

\begin{equation}
    \begin{aligned}
        s^2 &= 1 + h^2 + k^2;\\
        v &= h\sin(L)-k\cos(L);\\
        w &= 1 + f \cos(L) + g \sin(L);\\
    \end{aligned}
\end{equation}

\noindent and $\Delta_r$, $\Delta_t$ and $\Delta_n$ are perturbing accelerations in the radial, tangential, and normal directions of the spacecraft's LVLH frame. The LVLH and ECI frames are depicted in \autoref{fig:rsw reference frame}. 
The unit thrust vector in the ECI frame, \( \hat{\textbf{u}}_{\text{ECI}} \), is related to the unit thrust vector in the LVLH frame \( \hat{\textbf{u}}_{\text{MEE}} \) by \autoref{eq:LVLH frame unit vectors}.

\begin{subequations}
\label{eq:LVLH frame unit vectors}
\begin{equation}
\hat{\textbf{u}}_{\text{ECI}} = [\hat{\textbf{e}}_r \quad \hat{\textbf{e}}_\theta \quad \hat{\textbf{e}}_\phi]\hat{\textbf{u}}_{\text{MEE}}
\end{equation}
\begin{equation}
\hat{\textbf{e}}_r = \frac{\textbf{r}}{\norm{\textbf{r}}};
\quad
\hat{\textbf{e}}_\phi = \frac{\textbf{r} \times \textbf{v}}{\norm{\textbf{r} \times \textbf{v}}};
\quad
\hat{\textbf{e}}_\theta = \hat{\textbf{e}}_\phi \times \hat{\textbf{e}}_r
\end{equation}
\end{subequations}

Then, the thrust acceleration $\textbf{a}_{T}$ applied by the spacecraft in the LVLH frame is defined in \autoref{eq:thrust acceleration}, where $\hat{\textbf{u}}$ is the direction of application of thrust. The spacecraft's mass is propagated according to \autoref{eq:mass propagator}, assuming constant $I_{sp}$ through a single burn. 

\begin{equation}
\label{eq:thrust acceleration}
\textbf{a}_T = \frac{T}{m} \hat{\textbf{u}}
\end{equation}

\begin{equation}
\label{eq:mass propagator}
    \frac{\mathrm{d}m}{\mathrm{d}t}=\frac{T}{I_{sp}g_0};
\end{equation}

With regards to perturbations, the greatest impact for trajectory design in the case of near-Earth orbits comes from the Earth oblateness gravity potential distortion \cite{alfriend_chapter_2010}, characterized by the second zonal harmonic coefficient or $J_2$. The instantaneous acceleration components due to $J_2$ are included in the environmental model, which in the LVLH frame follow in \autoref{eq:J2 instantaneous acceleration}.

\begin{equation}
\label{eq:J2 instantaneous acceleration}
\begin{aligned}
\Delta_{J_{2,r}} &= -\frac{3\mu J_2 R_e^2}{2 r^4} \left(1 - \frac{12 v^2}{s^4} \right) \\
\Delta_{J_{2,\theta}} &= -\frac{12\mu J_2 R_e^2}{r^4} \left( \frac{v(h \cos L + k \sin L)}{s^4} \right) \\
\Delta_{J_{2,\phi}} &= -\frac{6\mu J_2 R_e^2}{r^4} \left( \frac{v(1 - h^2 - k^2)}{s^4} \right)
\end{aligned}
\end{equation}

The $J_2$ perturbing acceleration causes a secular perturbation on RAAN and AOP over a full orbit \cite{wakker_fundamentals_2015}, modelled by \autoref{eq:secular J2 perturbation}, where $n=\sqrt{\mu/a^3}$ is the mean motion of the orbiting body.

\begin{equation}
\label{eq:secular J2 perturbation}
\begin{aligned}
\frac{\mathrm{d}\Omega}{\mathrm{d}t}=&-\frac{3}{2}J_2\left(\frac{R_e}{p}\right)n\cos i;\\
\frac{\mathrm{d}\omega}{\mathrm{d}t}=&-\frac{3}{4}J_2\left(\frac{R_e}{p}\right)n(5\cos^2 i - 1));\\
\end{aligned}
\end{equation}

\section{Guidance Policies and Transfer Cost Estimation}
\label{sec: transfer}

As discussed in Section \ref{sec: mission}, OSSIE is constrained to multi-revolution impulsive manoeuvres. For the payload deployment mission scenario under consideration, OSSIE must achieve high accuracy SMA, inclination and phase convergence: RAAN targeting is not of interest to the current mission, as RAAN drift is very large for LEO orbits. This section presents the orbit guidance policies used to achieve insertion and the analytical models used to estimate their cost in $\Delta V$, fuel mass and time of flight. MHT manoeuvres are discussed in \autoref{subse:mht}, NIC manoeuvres in \autoref{subsec:nic}, and sequential MHT-NIC manoeuvres in \autoref{subse:seq}. Lastly, \autoref{subsec:J2 transfer} discusses the impact of the $J_2$ perturbation on the manoeuvres.

\subsection{Multiple Hohmann Transfer manoeuvres}
\label{subse:mht}

A Hohmann transfer is the optimal manoeuvre to transfer between two coplanar circular orbits of different radii. It is a two-impulse manoeuvre consisting of an initial burn, either raising apogee or lowering perigee (depending on objective), and a circularization burn when the apogee (or perigee) of the transfer manoeuvre is reached. For OSSIE, the $\Delta V$ required to perform a direct Hohmann transfer may not be achievable. Instead, a MHT approach is used. The MHT splits the departure and circularization burns into many consecutive insertion and circularization smaller burns, affordable by the maximum $\Delta V$ that can be injected at a time with the employed thrusters. 

\subsubsection{Required \texorpdfstring{$\Delta V$}{delta V}, fuel mass and number of burns}

The $\Delta V$ required by the MHT is equal to the one of a direct Hohmann Transfer achieving the same altitude change. The total magnitude can be computed as follows in \autoref{eq:mht cost}, where $\Delta V_d$ and $\Delta V_c$ stand for departure and circularization $\Delta V$, $V_0=\sqrt{\mu/r_0}$ is the orbital velocity at the departure orbit and $\xi=r_1/r_0$ is the ratio of target to departure orbit \cite{wakker_fundamentals_2015}.

\begin{subequations}
\label{eq:mht cost}
\begin{equation}
\Delta V_{\text{MHT}} = \Delta V_d + \Delta V_c
\end{equation}
\begin{equation}
    \Delta V_d = V_0 \abs{\sqrt{\frac{2n}{\xi+1}}-1}
\end{equation}
\begin{equation}
    \Delta V_c = V_0 \sqrt{\frac{1}{\xi}} \abs{\sqrt{\frac{2n}{\xi+1}}-1}
\end{equation}
\end{subequations}

The fuel mass required to perform the MHT is calculated as follows in \autoref{eq:fuel mass from delta v}, assuming constant $I_{sp}$ through the manoeuvre; the number $k$ of burns required to perform a manoeuvre is obtained as the ceil fraction of required fuel mass over mass flow $\dot{m}_f=T/(I_{sp}g_0)$ (see \autoref{eq:fuel mass from delta v}). The mass flow is assumed constant, as $T$ and $I_{sp}$ are assumed constant through the transfer.

\begin{equation}
\label{eq:fuel mass from delta v}
m_f = m_0 \left[1 - \exp\left(-\frac{\Delta V}{I_{sp}g_0}\right)\right]; \quad k = \ceil{\frac{m_f}{\dot{m}_f}}
\end{equation}

\subsubsection{Time Of Flight}

The total TOF of the MHT follows in \autoref{eq:mht tof}, where the burn frequency $f_{\text{burn}}$ is defined in \autoref{eq:burn freq}, $\olsi{P}$ stands for the average orbital period through the transfer defined in \autoref{eq:average tof mht}, and $k_d$ and $k_c$ are the number of departure and circularization burns respectively, defined in \autoref{eq:fuel mass from delta v}. 

\begin{equation}
\label{eq:mht tof}
\text{TOF}_\text{MHT}=\min(1, f_{\text{burn}})*(k_d + k_c)*\olsi{P}_{\text{MHT}}
\end{equation}

\begin{equation}
\label{eq:burn freq}
f_{\text{burn}} = (T_\text{burn}+T_\text{cooldown})^{-1}
\end{equation}

\begin{equation}
\label{eq:average tof mht}
\olsi{P}_{\text{MHT}}=\frac{4\pi}{5\sqrt{\mu}\left(r_{2}-r_{1}\right)}\left(r_{2}^{\frac{5}{2}}-r_{1}^{\ \frac{5}{2}}\right)
\end{equation}

\subsubsection{Phasing}

The spacecraft ends the MHT at true longitude $L_0+\pi$. To avoid phasing manoeuvres after reaching the required orbit, in order to acquire the target true longitude, the MHT start epoch is selected so that $L_0(t_0)=L_1(t_0+\text{TOF}_\text{MHT})-\pi$, where $TOF$ is the estimated time of flight. This correction is computed after the MHT calculation, thus $\text{TOF}_\text{MHT}$ is known.

\subsection{Nodal Inclination Change manoeuvres}
\label{subsec:nic}

An NIC manoeuvre modifies the inclination of an orbit without affecting its RAAN. Thrust is applied impulsively as the spacecraft crosses the ascending or descending node, such that the resulting velocity vector is that of the orbit with the desired inclination. A multiple NIC manoeuvre consists of splitting the impulsive $\Delta V$ that must be applied into multiple burns, up to twice per orbit, as the spacecraft crosses the orbit ascending and descending nodes.

\subsubsection{Required \texorpdfstring{$\Delta V$}{delta V}, fuel mass and number of burns}

The $\Delta V$ required for a multiple NIC is the same as for a single burn and follows in \autoref{eq:nic delta v}.  The fuel mass $m_f$ and number of burns $k_{\text{NIC}}$ required to perform the NIC is calculated using \autoref{eq:fuel mass from delta v} \cite{wakker_fundamentals_2015}.

\begin{equation}
\label{eq:nic delta v}
\Delta V_{\text{NIC}} = 2 V_0 \sin\left( \dfrac{\Delta i}{2} \right)
\end{equation}

\subsubsection{Time Of Flight}

The TOF of a multiple NIC is calculated with \autoref{eq:nic tof}, where $P=2\pi\sqrt{r^3/\mu}$ is the constant orbital period.

\begin{equation}
\label{eq:nic tof}
\text{TOF}_\text{NIC}=\min(2, f_{\text{burn}})*(k_{\text{NIC}})*P
\end{equation}

\subsection{Sequential MHT-NIC manoeuvres}
\label{subse:seq}

From \autoref{eq:nic delta v} it is of paramount importance to lower orbital velocity before performing an NIC. OSSIE will often have to reach targets with different semi-major axes and inclinations than its current ones: the most common case is orbit acquisition for payloads which require a particular orbital altitude to carry out their activity as well as a sun-synchronous orbit ---commonly the case for Earth observation and mapping satellites. \autoref{algo:seq} resolves sequential MHT and NIC manoeuvres by conducting the NIC when the semi-major axis is highest, such that the $\Delta V$ required for the NIC leg is minimized.

\begin{algorithm}
\caption{Sequential MHT-NIC Transfer}
\label{algo:seq}
\DontPrintSemicolon
\uIf(\tcp*[f]{Orbit raising}){$r_2 > r_1$}{
    \textbf{Step 1: MHT}\;
   Compute coast time to $L_0$\;
    Compute $\Delta V_{\text{MHT}}$\;
    Compute time of flight $t_{\text{MHT}}$\;
    \textbf{Step 2: NIC}\;
    Compute coast time to closest node\;
    Compute $\Delta V_{\text{NIC}}$\;
    Compute time of flight $t_{\text{NIC}}$\;
}
\Else(\tcp*[f]{Orbit lowering}){
    \textbf{Step 1: NIC}\;
    Compute coast time to closest node\;
    Compute $\Delta V_{\text{NIC}}$\;
    Compute time of flight $t_{\text{NIC}}$\;
    \textbf{Step 2: MHT}\;
    Compute coast time to $L_0$\;
    Compute $\Delta V_{\text{MHT}}$\;
    Compute time of flight $t_{\text{MHT}}$\;
}
\end{algorithm}

\subsection{\texorpdfstring{$J_2$}{J2} perturbation}
\label{subsec:J2 transfer}

RAAN and AOP drift according to \autoref{eq:secular J2 perturbation} for the duration of the manoeuvre. In the case of the MHT this only impacts the phasing coasting leg due to the drift of AOP. The NIC manoeuvre is unaffected.

\section{Trajectory Optimization}
\label{sec:trajopt}

This section presents the optimization framework proposed for solving the problem introduced in \autoref{sec: mission} subject to dynamics of \autoref{sec:EnvironmentModel} and manoeuvres as described in \autoref{sec: transfer}.
In \autoref{subsec:solver arch} we present a modular solver architecture based on integer-nonlinear decomposition and heuristic combinatorial optimization, capable of integrating diverse trajectory estimation methods, heuristic optimization algorithms, and trajectory refinement procedures. \autoref{subsec:solver combinatorial} discusses the combinatorial optimization components of the solver. \autoref{subsec:generation} discusses trajectory generation, and \autoref{subsec:scp} presents the SCP trajectory re-optimization block and specific tailoring to the OSSIE mission design case.

\subsection{Solver Architecture}
\label{subsec:solver arch}

\begin{figure*}
    \centering
    \includegraphics[width=\linewidth]{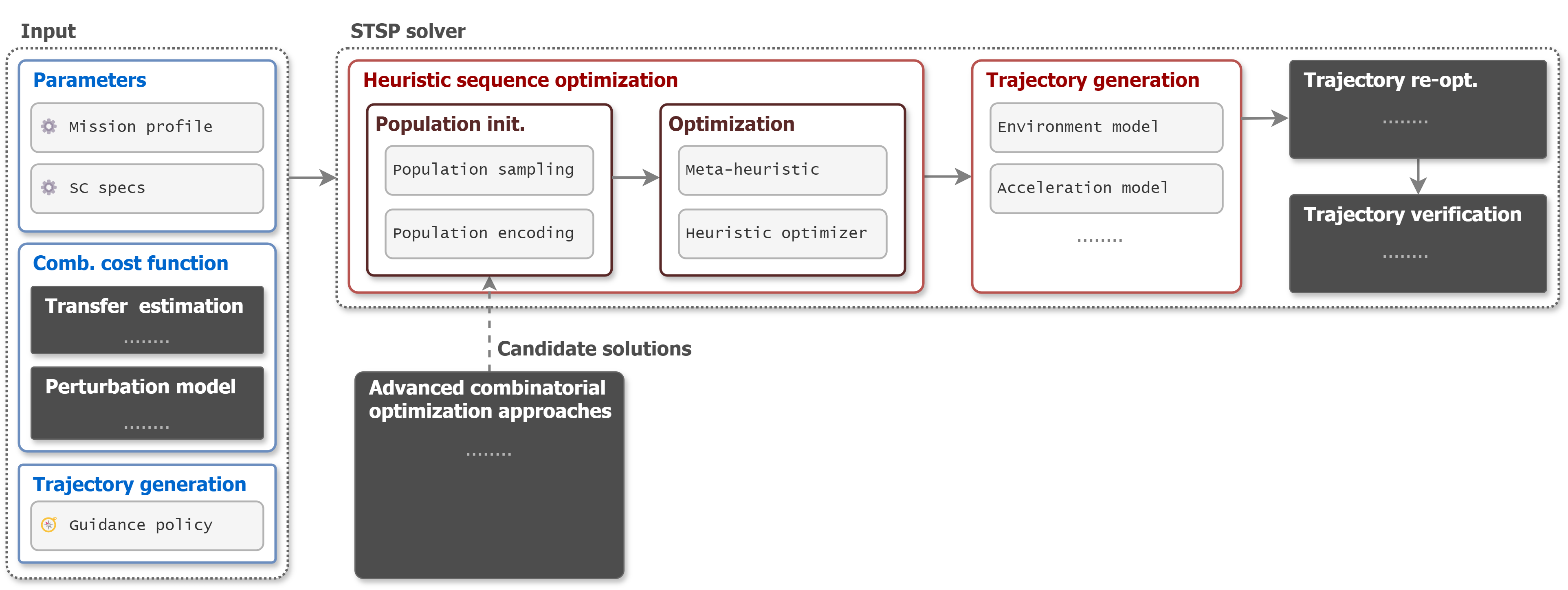}
    \caption{STSP solver architecture. In black: complex components (integrated using standardized interfaces) the internal structure of which is out of the scope of this diagram.}
    \label{fig:solver arch}
\end{figure*}

Our aim is to design a highly adaptable solver capable of generating multi-rendezvous trajectories for spacecraft guidance with strict feasibility guarantees, and to use it to design trajectories for the OSSIE OTV. We achieve this using a 3-stage solver architecture (\autoref{fig:solver arch}) consisting of target sequence optimization, trajectory generation and re-optimization, and trajectory verification. Each component is implemented modularly using standardized interfaces, resulting in a framework that is highly adaptable to new mission design scenarios, while benefiting from high component reusability.

\subsection{Sequence Optimization Problem}
\label{subsec:solver combinatorial}

We make use of heuristic, population-based combinatorial optimization to optimize target sequences. Knowledge of near-optimal solutions, obtained by any means, is leveraged by initializing populations using distance-based permutation sampling. Later in \autoref{sec:rl} we demonstrate this versatility by integrating an attention-based STSP routing model trained using RL to greatly improve convergence to the global optimum. 

\subsubsection{Population Sampling and Encoding}
\label{subsubsec:sampling}

High quality population sampling is crucial for population-based global combinatorial optimization \cite{mitchell_sampling_2022}. We make use of uniform permutation sampling to cover the search space as widely as possible. The combinatorial optimization space is vast however, and advanced methods and hand-crafted heuristics are often used to determine near-optimal solutions prior to heuristic optimization \cite{izzo_evolving_2015,petropoulos_gtoc9_2017}. When these are available we make use of distance-based permutation sampling to spawn initial populations in the vicinity of candidate solutions, greatly helping the search space exploration process. We find that a combination of both approaches is optimal to balance exploration and exploitation of the search space.

\paragraph{Uniform Permutation Sampling} We uniformly sample the permutation group $\mathfrak{S}^n$ by drawing Sobol points in the $[0,1)^n$ hypercube and applying an \texttt{argsort} operation. This algorithm was found to yield superior samples than the Fisher-Yates shuffle \cite{eberl_fisher-yates_2016} and Knuth's algorithm \cite{mitchell_sampling_2022}, while being orders of magnitude faster than advanced approaches such as Sobol Permutations \cite{mitchell_sampling_2022}.

\paragraph{Distance-based Permutation Sampling}
The Mallows model \cite{diaconis_group_1988} (and related methods), first introduced in 1957 \cite{mallows_non-null_1957}, is the most relevant distance-based statistical model for permutations \cite{irurozki_sampling_2014}. Permutations are exponentially less probable as they stray further from a central permutation $\sigma_0$. Refer to Irurozki, 2014 \cite{irurozki_sampling_2014} for a comprehensive definition of the model.

\paragraph{Population Encoding} Random keys permutation encoding \cite{bean_genetic_1994} is selected for its versatility for optimizing complex discrete optimization problems across various domains \cite{kromer_novel_2022,londe_biased_2024}. Permutations are encoded using vectors of continuous values, or random keys, in the $[0, 1)$ range. A permutation $\sigma\in\mathfrak{S}^n$ is encoded by generating a vector of random keys $\mathbf{x}\in[0,1)^n$, sorting it, and permuting it by $\sigma$. Decoding is done by applying an \texttt{argsort} operation to $\mathbf{x}$. 

\subsubsection{Combinatorial optimization}
\label{subsubsec:comb}

The generalized STSP to be solved is described in \autoref{eq:stsp problem}, where $T$ is the set of all targets and $c_\pi$ is the cumulative cost function of a tour. The cumulative cost function sequentially estimates transfer cost and time of flight and propagates the environment according to a perturbation model (see inputs in \autoref{fig:solver arch} for reference). Transfer cost estimation and environment propagation are implemented modularly, resulting in a highly flexible modelling process and high module reusability. In the case of OSSIE this is the $J_2$ secular perturbation model in \autoref{eq:secular J2 perturbation}. Departure state $h$ and a decommissioning orbit $d$ are implemented as optional search space constraints. 

\begin{mini}
  {\pi}{\sum_{k=0}^{n-1} c_{\pi(k), \pi(k+1)}}{\label{eq:stsp problem}}{}
  \addConstraint{\pi(0)}{= h}
  \addConstraint{\pi(n)}{= 
    \begin{cases}
      h & \text{(Hamiltonian cycle)} \\
      d & \text{(Decom.)}
    \end{cases}}
  \addConstraint{\{\pi(k)\}_{k=1}^{n-1},}
                {= T \setminus \{h\}}
  \addConstraint{\pi(k)}
                {\in T \cup \{d\},\quad}
                {k \in 0,1,\dots, n}
\end{mini}

The combinatorial optimization component is implemented using the \texttt{pygmo} \footnote{\url{https://esa.github.io/pygmo2/overview.html}} parallel multi-objective global optimization library, which is based on the underlying PAGMO C++ library \cite{biscani_parallel_2020}. This choice allows us to quickly evaluate the performance of a wide variety of global and local optimization algorithms, and design optimal solvers for particular mission design scenarios such as that of OSSIE. 

\paragraph{Meta-heuristic}
Combinatorial optimization meta-heuristics are sophisticated algorithmic frameworks designed to efficiently explore large and complex discrete search spaces to identify optimal or near-optimal solutions for combinatorial problems. 
We make use of the Archipelago meta-heuristic implemented in \texttt{pygmo}. The Archipelago meta-heuristic is based on the concurrent evolution of sub-populations, or "islands," starting from distinct initial populations, and possibly using distinct evolutionary strategies. Periodic migrations of individuals between islands promote diversity and prevent premature convergence, enhancing the algorithm's ability to escape local optima and explore diverse regions of the search space. This parallel and cooperative framework not only accelerates the optimization process but also improves the robustness and quality of the solutions obtained.

\paragraph{Heuristic Optimization Algorithm} Optimizer selection is conducted by trading off the performance of all global heuristic optimizers available in \texttt{pygmo} (refer to the \texttt{pygmo} capabilities page\footnote{\url{https://esa.github.io/pygmo2/overview.html}}) on the STSP variant at hand. Critically, optimizers of the family of Evolutionary Strategies depend on internal sampling processes and are thus not sensitive to initial populations: in general terms other optimizers are preferable if candidate solutions can be leveraged (this is not the case for OSSIE, as will be shown in \autoref{sec:results}).

\subsection{Trajectory Generation}
\label{subsec:generation}

Trajectories are generated by propagating the state of the spacecraft under a guidance policy. The guidance policy is implemented as a state machine that determines when and how to apply thrust through the manoeuvre, following the guidance policy API of the Tudat Space\footnote{\url{https://docs.tudat.space/en/latest/}} astrodynamics library \cite{dirkx_open-source_2022}. This provides ample flexibility for the implementation of the simulator itself. In the case of OSSIE we generate trajectories using a simple analytical propagator, considering impulsive orbital changes and secular $J_2$ perturbations (see \autoref{subsec:J2 transfer}). In more complex scenarios we make use of Tudat to implement and refine the simulator used to generate spacecraft trajectories.

\subsection{Trajectory re-optimization with Sequential Convex Programming}
\label{subsec:scp}

The final block of the optimization chain is in charge of adapting the ideal transfer maneuvers to feasible spacecraft trajectories. To account for actuation constraints, by introducing the state vector:  
\begin{equation}
\begin{aligned}
  & x = [p, f, g, h, k, L, m]^T   \\
\end{aligned}
\end{equation}
as well as the control vector:
\begin{equation}
\begin{aligned}
  & u = [u_{r}, u_{\theta}, u_{\phi}]^T   \\
\end{aligned}
\end{equation}
the following OCP is formulated for each transfer arc: 
\begin{equation}\label{eq:SCP_OCP}
\begin{aligned}
\min_{X\in \mathbb{R}^{n_x}, U \in \mathbb{R}^{n_u}} \quad & \frac{1}{2}(x_N-x_{{ref}})^{T}P(x_N-x_{{ref}})\\ &+ \frac{1}{2} U^{T}RU\\
\textrm{s.t.} \quad & x_{i+1} = f(x_i, u_i)\quad i \in \{0,\dots, N-1\} \\
  & \Vert u_i \Vert \leq T_{\text{max}_i} \quad i \in \{0,\dots, N-1\}   \\
    &      x_0 = \hat{x_0} \\
\end{aligned}
\end{equation}
where $N$ is the length of the optimization horizon, i.e., the number of stages; $n_x = 7(N+1)$ and $n_u=3N$ the total number of state and control optimization variables, respectively; so that ${X} = \mathrm{col}({x}_0,\cdots,{x}_N)$ and ${U} = \mathrm{col}({u}_0,\cdots,{u}_{N-1})$; $f(x_i,u_i)$ the discretized non-linear orbital dynamics, $\hat{x_0}$ the initial state and being $x_{ref}$ the reference trajectory, i.e., the optimized arc obtained from the combinatorial problem.

The Euclidean norm of the control input $||\cdot||$ is constrained to a maximum thrust $T_{\text{max}_i}$, mission and potentially time-dependent. The penalty of the thrust action is weighted by $R\in\mathbb{R}^{n_u\times n_u} $, $R\succcurlyeq0$, against the final error $(x_N-x_{ref})$, weighted by matrix $P\in\mathbb{R}^{7\times 7}$, $P\succcurlyeq0$. It is highlighted that here the penalty over the final state error is a relaxation of the hard constraint $x(t_N) = x_N$ with $x_N$ the final state of each arc of trajectory from the combinatorial problem, to avoid unfeasibilities when adapting to spacecraft dynamics and constraints.   
The TOF between $\hat{x_0}$ and $x_{N}$ is then used to adapt the optimization horizon. Note that the number of arcs simultaneously optimized determines the authority given to this layer of the framework to modify the transfers, being the case with a single arc focused on replicating the Hohmann arc, autonomously adapting it to the spacecraft constraints.

Finally, note that the OCP in \autoref{eq:SCP_OCP} does not include attitude constraints, e.g., for guaranteeing that the thrust vector can be oriented in time to perform the required impulses. Certain smoothness of the attitude can be argued in the implicit effect of the control cost regularization, yet the formulation is currently under development for inclusion of explicit attitude constraints.

\subsubsection{Solution with SCP}\label{sec:SCPalgo}
The resultant generic OCP in \autoref{eq:SCP_OCP} is convex except for the non-linear dynamics implicit in $f(x_i,u_i)$. To solve it, an SCP solver is used from the SOTB\footnote{SOTB is an independent in-house toolbox developed in \texttt{MATLAB} devoted for real-time embedded applications. 
Within available solvers, SCP algorithms focused on nonlinear numerical optimal control powered by Interior Point Method (IPM) solvers are ideal for the developed application.} library. 

The algorithm implemented follows a classical SCP logic, solving  guess-based convexified iterations of the original problem until
a convergence criterion is matched. Dynamic trust regions are employed to control the iteration progress and avoid divergence or solution chattering due to non-linear effects. We refer to \cite{ramirez} for further details on the solver.

To reduce errors arising from numerical discretization and propagation during the iterations of the optimization solver, the same MEEs formulation of Section \ref{sec:EnvironmentModel}
is adopted in the SCP with a proper discretization step for the horizon length and integrator selection. Furthermore, the optimization variables and the dynamics model have been normalized, i.e., the state variables have been scaled according to the MEE parametrization of the latter point of each optimization horizon, and the control input sequence has been scaled according to the maximum value of the thrust. This normalization moves the optimum region to similar orders of magnitude in states and control, providing a smoother behavior of the SCP internal IPM and smaller linearization errors, aside from easing the algorithm tuning.

\subsubsection{OSSIE actuation constraints}

\begin{figure}[htp]
    \centering
    \includegraphics[width=
    \linewidth]{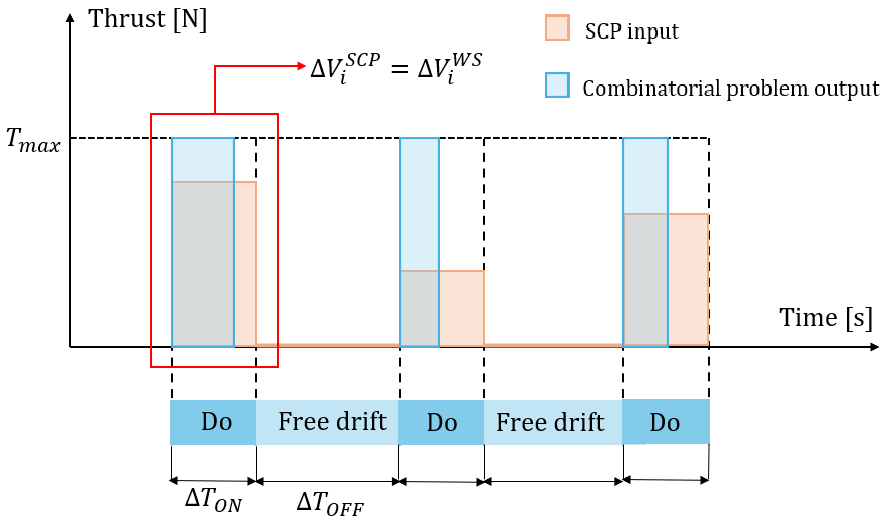}
    \caption{On-time constraints scheme and warm-starting procedure}
    \label{fig:Tmax}
\end{figure}

The propulsion system of the OSSIE OTV is not capable of continuous firing for longer than $\Delta T_{\text{on}}$ seconds, requiring a minimum off-time for cool-down after a burn. To comply with the maximum firing time of the OSSIE thruster actuators, $T_{\text{max}_i}$ shall be set so that after the maximum firing time the constraint is set to zero until the minimum off-time is reached. To simplify the problem and avoid introducing the time as a constrained optimization variable, the input TOF between consecutive  $\Delta V$ from the combinatorial problem is used to compute $T_{\text{max}_i}$ as sketched in Fig.~\ref{fig:Tmax}, i.e., forcing $T_{\text{max}_i}=0$ for transfer periods which are already larger than the minimum off-time. 

Note that this decision enforces the actuation windows based on the combinatorial solution and Hohmann transfers.  For multiple arcs optimization, this might create a source of suboptimality. Nonetheless, for the single arc optimization, this does not generate a problem because maintaining the TOF is an objective thus the action window coincides with the perigee or apogee, i.e., the most efficient times to act. This simplification results in a pragmatic solution because few simultaneous arcs optimization is sufficient in the considered mission. 

\subsubsection{Multi-impulsive warm start}
Warm starting can significantly reduce the number of required iterations in an SCP algorithm. For the mission case, an efficient initialization is created with the solution of the combinatorial problem. The departure and complete $\Delta V$ time sequence are used to automatically adapt the discretization scheme of the trajectory arc(s) optimization and on-time constraints, based on the number of consecutive arcs to be simultaneously optimized (parameter). Then, $\Delta V$s are translated to forces realizing equivalent total pulse but uniformly distributed in $\Delta T_{\text{on}}$. Fig.~\ref{fig:Tmax} illustrates the warm start creation process. This approach generally situates the initial guess close to the optimum, significantly reducing the required convexification iterations. 

\section{Reinforcement Learning for Combinatorial Optimization}
\label{sec:rl}

ML approaches for spacecraft trajectory design have seen a surge of interest in recent years \cite{izzo_survey_2019}, with strong results achieved both for trajectory cost estimation \cite{li_deep_2020} and spacecraft guidance \cite{izzo_interplanetary_2019}. NCO uses DNNs to automate the problem-solving process, mostly under the RL paradigm, as supervised learning is often infeasiblefor large or theoretically hard problems. NCO offers the attractive prospect of alleviating the scaling issues of exact approaches, while removing the need for handcrafted heuristics, which often require significant domain-specific adjustments \cite{berto_rl4co_2024}. NCO has shown promising performance on various combinatorial optimization problems \cite{berto_rl4co_2024},
especially when coupled with advanced policy search procedures \cite{francois_how_2019}.

We make use of \texttt{RL4CO}\footnote{\url{https://rl4.co}} to implement and train the STSP routing policy. \texttt{RL4CO} is a benchmark library for NCO with standardized, modular, and highly performant implementations of various environments, policies and RL algorithms, covering the entire NCO pipeline \cite{berto_rl4co_2024}.

\subsection{Environment}

The space environment is modelled in MEEs. The translational state is extended with the mass of the vehicle and targets. The resulting 7-dimensional state space is embedded using a feedforward NN into a 128-dimensional space in which the policy operates. The global state is propagated through the sequential decision-making process. Environment propagation is implemented by means of a transfer TOF estimation method and perturbation model: in the case of OSSIE these are the sequential MHT-NIC trajectory estimator, and secular $J_2$ perturbation model defined in \autoref{sec: transfer}.

\subsection{Policy}

We make use of an autoregressive attention model\footnote{\href{https://rl4.co/docs/content/api/zoo/constructive_ar/\#models.zoo.am.policy.AttentionModelPolicy}{\texttt{https://rl4.co/docs/.../AttentionModelPolicy}}} first introduced by Kool et al. \cite{kool_attention_2019}, which encodes the input graph using a GAT, and decodes the solution using a Pointer Network. This policy trained using the REINFORCE RL algorithm has been demonstrated to perform considerably better than other learned heuristics \cite{kool_attention_2019}, as well as to have a highly efficient learning component \cite{francois_how_2019}.

\subsection{Reinforcement Learning Algorithms}

We make use of three RL algorithms implemented in \texttt{RL4CO} to train the policy: REINFORCE, Advantage Actor-Critic, and Proximal Policy Optimization.

\paragraph{Stochastic Policy Gradient}

The REINFORCE algorithm, introduced by Williams \cite{williams_simple_1992}, is a stochastic policy gradient method which uses Monte Carlo sampling to estimate policy gradient. By generating complete trajectories and using the returns from these trajectories, REINFORCE provides an unbiased estimate of the gradient. It updates the policy parameters by maximizing the expected reward through gradient ascent. However, REINFORCE often suffers from high variance in gradient estimates, which can hinder convergence.

\paragraph{Advantage Actor-Critic}

A2C methods enhance REINFORCE by incorporating a baseline to reduce the variance of gradient estimates. Introduced by Konda and Tsitsiklis \cite{konda_actor-critic_1999} and popularized in the asynchronous framework by Mnih et al. \cite{mnih_asynchronous_2016}, A2C simultaneously learns a policy (actor) and a value function (critic). The critic estimates the value function, which serves as a baseline to compute the advantage, thereby stabilizing and accelerating training.

\begin{table*}[t]
\centering
\caption{Top: OSSIE insertion and decommissioning orbits. Bottom: statistical model describing expected payload Keplerian states. In commercial operations the target state of each payload is specified by the client. SSO stands for SSO inclination variance.}
\label{tab: OSSIE mission statistical model}
\begin{adjustbox}{max width=\linewidth}
\begin{tabular}{lllllllll}
\thickhline
                         & Orbit        & $a$ [km] & $e$ [-]     & $i$ [deg] & $\Omega$ [deg] & $\omega$ [deg] & $\theta$ [deg] & Mass [kg] \\ \cline{2-9}
\multirow{2}{*}{OSSIE}   & Insertion       & 500      & 0           & 97        & 158            & 0              & 0           &   \\
                         & Decommissioning & 250      & 0           & -         & -              & -              & -           &   \\ \hline
\multirow{3}{*}{Payload} & Nominal         & 500      & 0           & 97        & 158            & 0              & 0             & Variable \\
                         & Spread          & 50       & 0           & SSO       & 180               & 180            & 180         & 15\%   \\
                         & Distribution    & Uniform  & Exponential & Uniform   & Uniform        & Uniform        & Uniform  & Exponential\\
\thickhline
\end{tabular}
\end{adjustbox}
\end{table*}

\paragraph{Proximal Policy Optimization}

PPO, introduced by Schulman et al. \cite{schulman_proximal_2017}, addresses some of the limitations of A2C by ensuring that policy updates do not deviate excessively from the current policy. PPO employs a clipped objective function to constrain the policy updates, balancing exploration and exploitation more effectively. This leads to more stable and reliable training, making PPO a popular choice for various RL applications, including combinatorial optimization tasks.

\subsection{Policy Search}

Policy search strategies determine how actions are selected based on the learned policy, balancing exploration and exploitation to find optimal or near-optimal solutions. Policy search is fundamental for the performance of NCO algorithms \cite{francois_how_2019}. We consider three policy search strategies implemented in \texttt{RL4CO}: greedy search, sampling and beam search.

\section{Results}
\label{sec:results}

This section presents empirical results obtained for the OSSIE OTV trajectory optimization problem. \autoref{subsec:mission modelling} describes the mission scenarios for OSSIE and presents a statistical model used to randomly generate realistic mission scenarios. In \autoref{subsec: combinatorial results} we report and analyze heuristic optimization performance. In \autoref{subsec:rl results} we analyze the performance of the RL Attention-based routing policy for this STSP variant. In \autoref{subsec:mission analysis} presents a study of OSSIE's mission design envelope based on a Monte Carlo campaign spanning all feasible mission modes. \autoref{subsec: results scp} presents the trajectory re-optimization results obtained using Sequential Convex Programming. Lastly, \autoref{subsec: verification results} discusses the verification of the generated trajectories on the OSSIE Functional Engineering Simulator developed at SENER.

\subsection{Mission Modelling}
\label{subsec:mission modelling}

\autoref{tab: OSSIE mission statistical model} describes the nominal mission scenario for OSSIE and the statistical model used to generate the target Keplerian states for a given payload. 
OSSIE payloads are highly likely to be destined for SSO orbits with altitude priority, but not necessarily. A worst case scenario is assumed, uniformly sampling payload inclinations in the entire SSO inclination range, defined as the range from $i_{\text{SSO}}(a_{\text{min}})$ to $i_{\text{SSO}}(a_{\text{max}})$, where $i_\text{SSO}(a)$ is the inclination required to achieve an SSO orbit (RAAN drift of 360\textdegree per year, see \autoref{eq:secular J2 perturbation}) at a given $a$. 
Payload masses are sampled from an exponential distribution, assuming a worst case scenario where payload masses are never below their nominal value: 6 kg for CubeSats, 1.5 kg for PocketQubes and 25 kg for small satellites. Payloads may be released individually or at once. 

OSSIE payloads may be released individually or bundled with other payloads. This variable is both client-dependent and so highly unpredictable, and greatly impactful for mission cost as it determines the number of transfers that must be carried out. We model this by uniformly sampling number of bundles between 2 (all payloads in two bundles) and the total number of payloads, which is 13 for OSSIE (every payload launched to a distinct target orbit); payloads are assigned to each bundle uniformly at random.

\begin{table*}[t]
\caption{Policy test performance and training time for the three RL algorithms considered. BS: Beam Search.}
\label{tab:rl results}
\centering
\begin{adjustbox}{max width=\linewidth}
\begin{tabular}{llllllllll}
\thickhline
RL algorithm        & \multicolumn{3}{c}{REINFORCE}             & \multicolumn{3}{c}{A2C}                & \multicolumn{3}{c}{PPO}                \\ \hline
Search strategy     & Greedy & Stoch. & BS    & Greedy & Stoch. & BS & Greedy & Stoch. & BS \\ \hline
Fuel mass [kg]      & 21.65           & 22.58           & \textbf{20.67} & 24.31           & 25.51           & 22.93       & 22.25           & 22.98           & 21.14       \\
Optimality gap [\%]  & 7.86            & 12.53           & \textbf{3.02}  & 21.17           & 27.15           & 14.28       & 10.86           & 14.52           & 5.35        \\
Training time [min] & 15.1            & -               & -              & \textbf{9.8}    & -               & -           & 30.6            & -               & -           \\
\thickhline
\end{tabular}
\end{adjustbox}
\end{table*}

\begin{figure}[h]
    \centering
    \includegraphics[width=\linewidth]{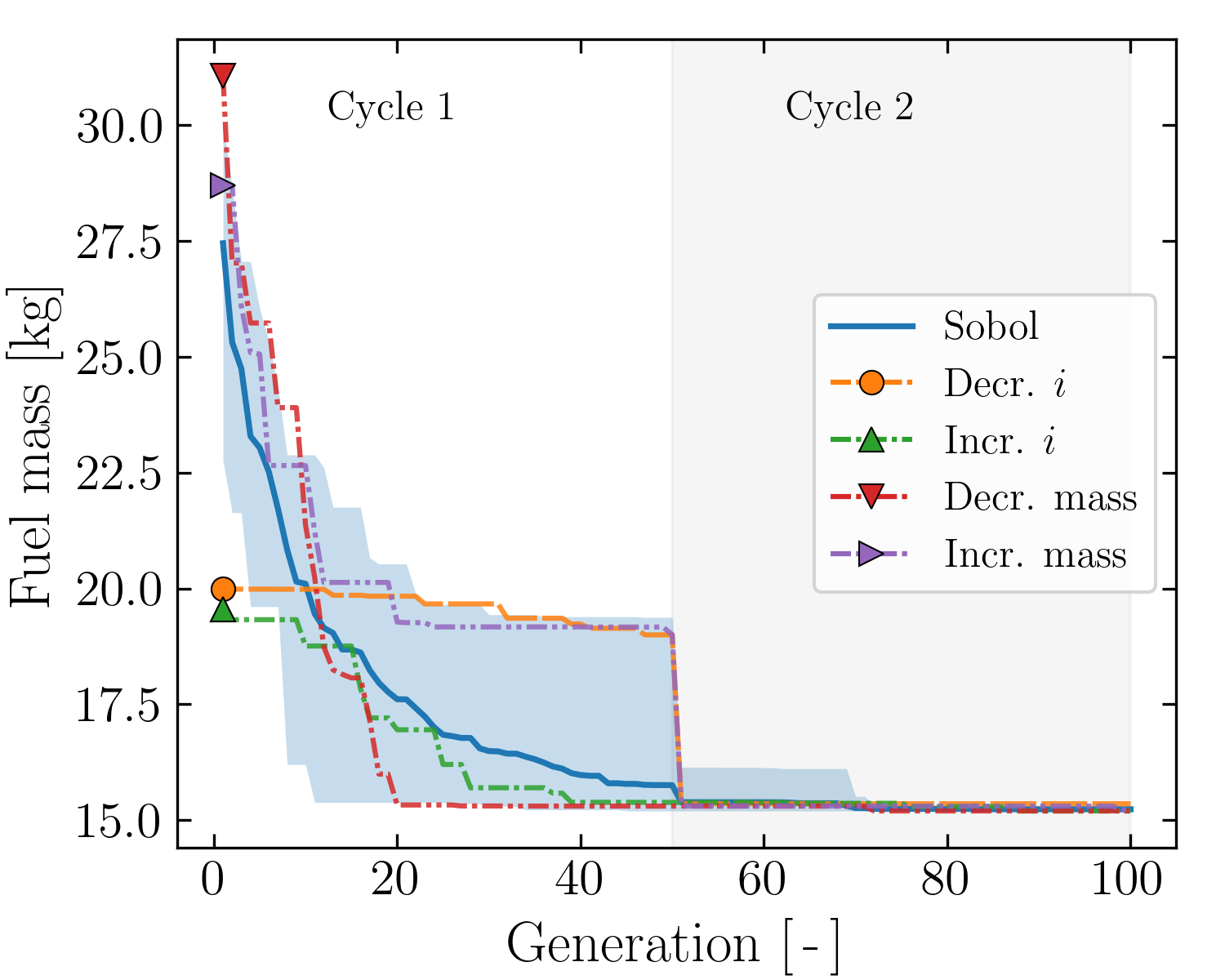}
    \caption{Best fuel mass achieved as a function of generation (13-transfer mission, GPSO). Shaded regions: evolutionary cycles.}
    \label{fig:gpso learning curve w initial guesses}
\end{figure}

\begin{figure*}[t]
    \begin{minipage}[t]{0.485\linewidth}
        \centering
        \includegraphics[width=\linewidth]{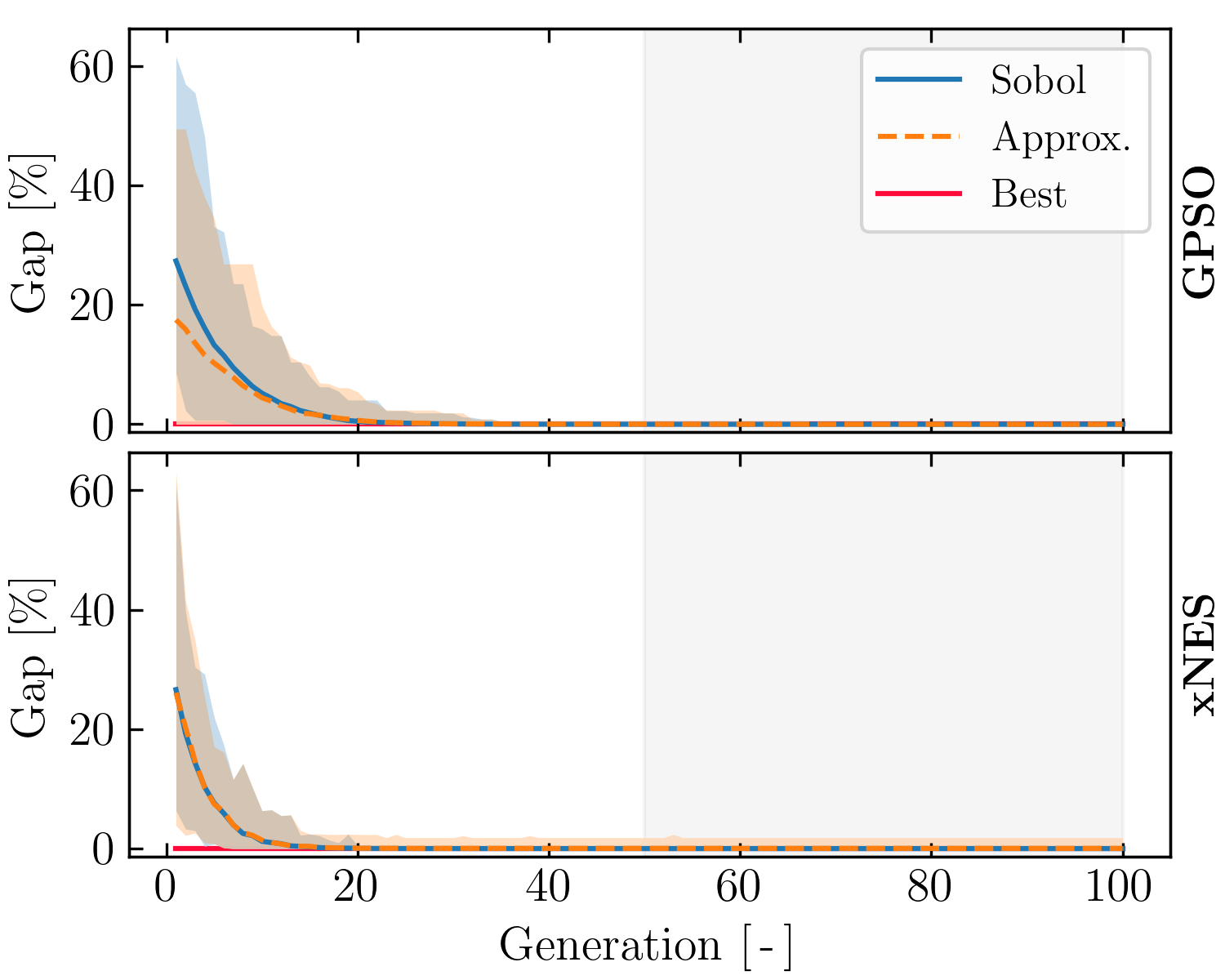}
        \caption{Optimization of 100 randomized 13-transfer OSSIE mission scenarios using GPSO and xNES. Best: on average 26.72 [kg] of fuel mass spent with a standard deviation of 2.25 [kg]. Shaded regions: evolutionary cycles.}
        \label{fig:learning curve monte carlo}
    \end{minipage}\hfill%
    \begin{minipage}[t]{0.485\linewidth}
        \centering
        \includegraphics[width=\linewidth]{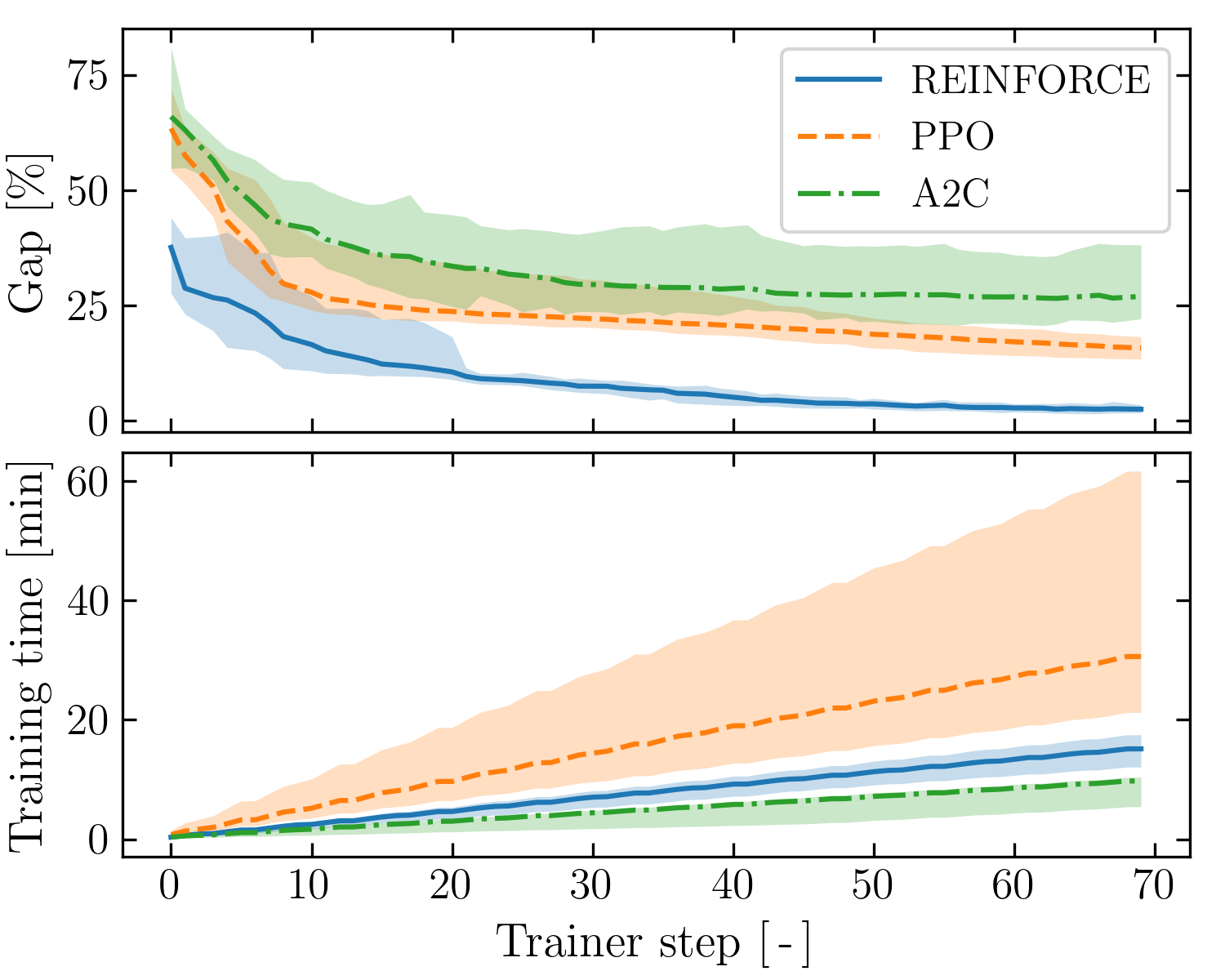}
        \caption{Training validation reward curve (expressed as a mean optimality gap) with REINFORCE, A2C and PPO.}
        \label{fig:RL model reward}
    \end{minipage}
\end{figure*}

\begin{figure*}[t]
    \centering
    \includegraphics[width=\linewidth]{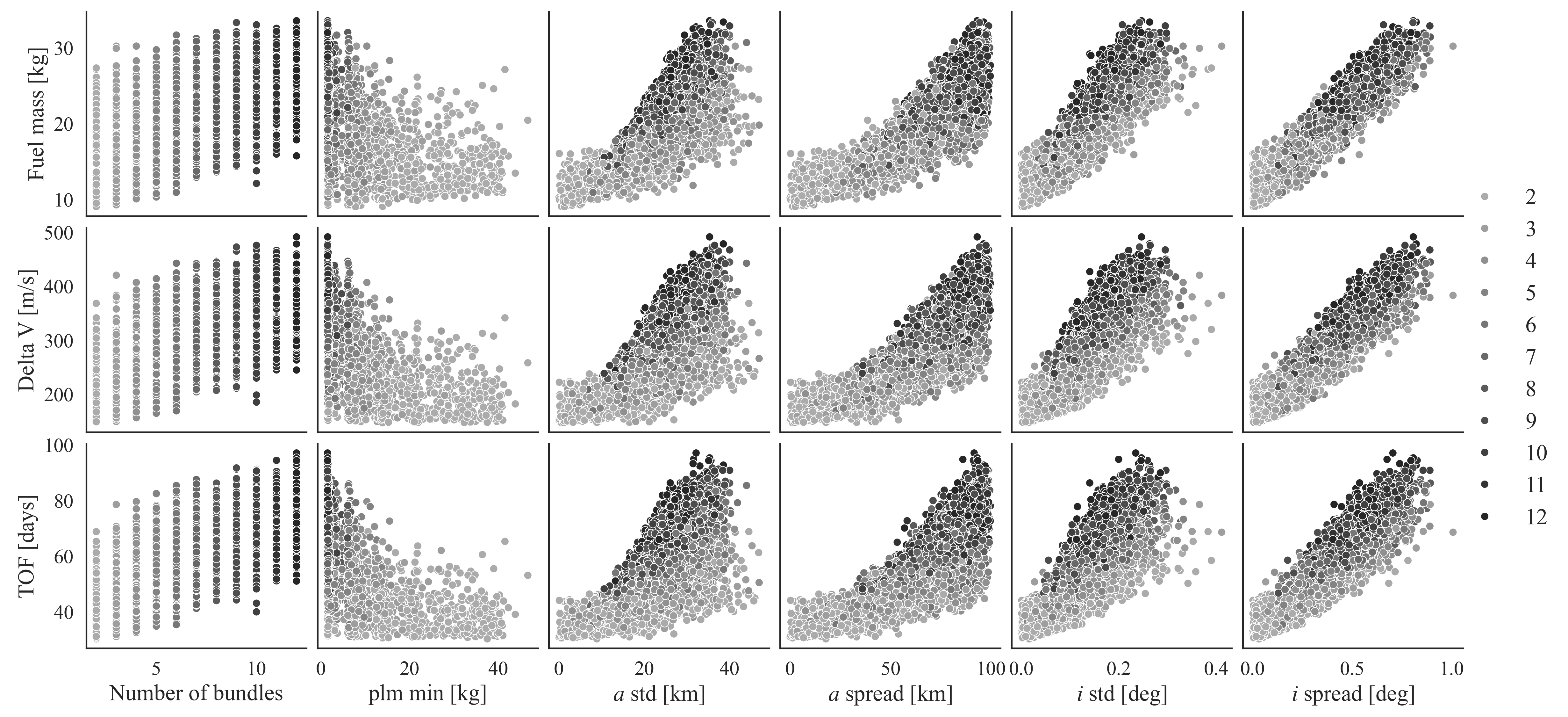}
    \caption{Cost of 5000 optimized mission scenarios. Fuel mass, $\Delta V$ and TOF are plotted against, from left to right: number of bundles, minimum payload mass, semi-major axis standard deviation and range, and inclination standard deviation and range. Observe the strong correlations between number of bundles and cost, and between inclination spread and cost.}
    \label{fig:pairplot}
\end{figure*}

\subsection{Heuristic Combinatorial Optimization}
\label{subsec: combinatorial results}

We find that GPSO\footnote{\texttt{\href{https://esa.github.io/pygmo2/algorithms.html\#pygmo.pso_gen}{https://esa.github.io/pygmo2/.../pso\_gen}}} and xNES\footnote{\texttt{\href{https://esa.github.io/pygmo2/algorithms.html\#pygmo.xnes}{https://esa.github.io/pygmo2/.../xnes}}} perform best in the case of OSSIE. \autoref{fig:gpso learning curve w initial guesses} shows the evolutionary curve of a realistic 13-transfer mission scenario for OSSIE using GPSO, which is sensitive to initial populations. \autoref{fig:learning curve monte carlo} shows the evolutionary curve of 100 randomized 13-transfer mission scenarios. 

Four hand-crafted heuristics are used to provide candidate solutions: ascending and descending inclination and payload mass walks.
The increasing inclination walk is a fairly good heuristic in \autoref{fig:gpso learning curve w initial guesses}, but this fails to replicate over more cases, with optimizations leveraging the four heuristics performing very similarly to those starting with uniformly sampled populations in \autoref{fig:learning curve monte carlo}. xNES outperforms GPSO in this case, exhibiting more uniform convergence. This dynamic only increases with problem size: improved heuristics are highly desirable, and more so as problem size increases, as is likely to be the case through the operational life of OSSIE.

\subsection{Reinforcement Learning}
\label{subsec:rl results}

We train the Attention policy on 100,000 10-transfer mission scenarios using a batch size of 5096, for 50 epochs. The policy is tested on 1000 mission scenarios. Near-optimal solutions for all scenarios are obtained a priori using heuristic optimization, and used as a benchmark to calculate optimality gaps. Training is repeated 10 times with each algorithm using different random seeds. 

\autoref{tab:rl results} reports mean policy training times and performances obtained with REINFORCE, A2C and PPO. \autoref{fig:RL model reward} shows the corresponding learning curves. REINFORCE yields the best training out of the three RL algorithms, with a mean optimality with respect to the heuristic solutions of 3.02\% using Beam Search. As expected \cite{francois_how_2019} Beam Search improves the final result compared to the greedy and sampling searches. The policy's performance is superior to that of the hand-crafted heuristics considered for OSSIE (\autoref{fig:learning curve monte carlo}). 

Having a reliably performant learned heuristic is considerably beneficial for combinatorial optimization (see \autoref{fig:gpso learning curve w initial guesses}); in the sections that follow we use GACO with initial populations determined by the policy to optimize OSSIE mission scenarios.

\subsection{Mission Analysis}
\label{subsec:mission analysis}

We proceed to perform a Monte Carlo analysis covering all feasible mission scenarios based on the nominal payload list of OSSIE of 8 CubeSats, 4 PocketQubes and 1 small satellite. 5000 mission scenarios are solved and reported. 

\begin{figure}
    \centering
    \includegraphics[width=\linewidth]{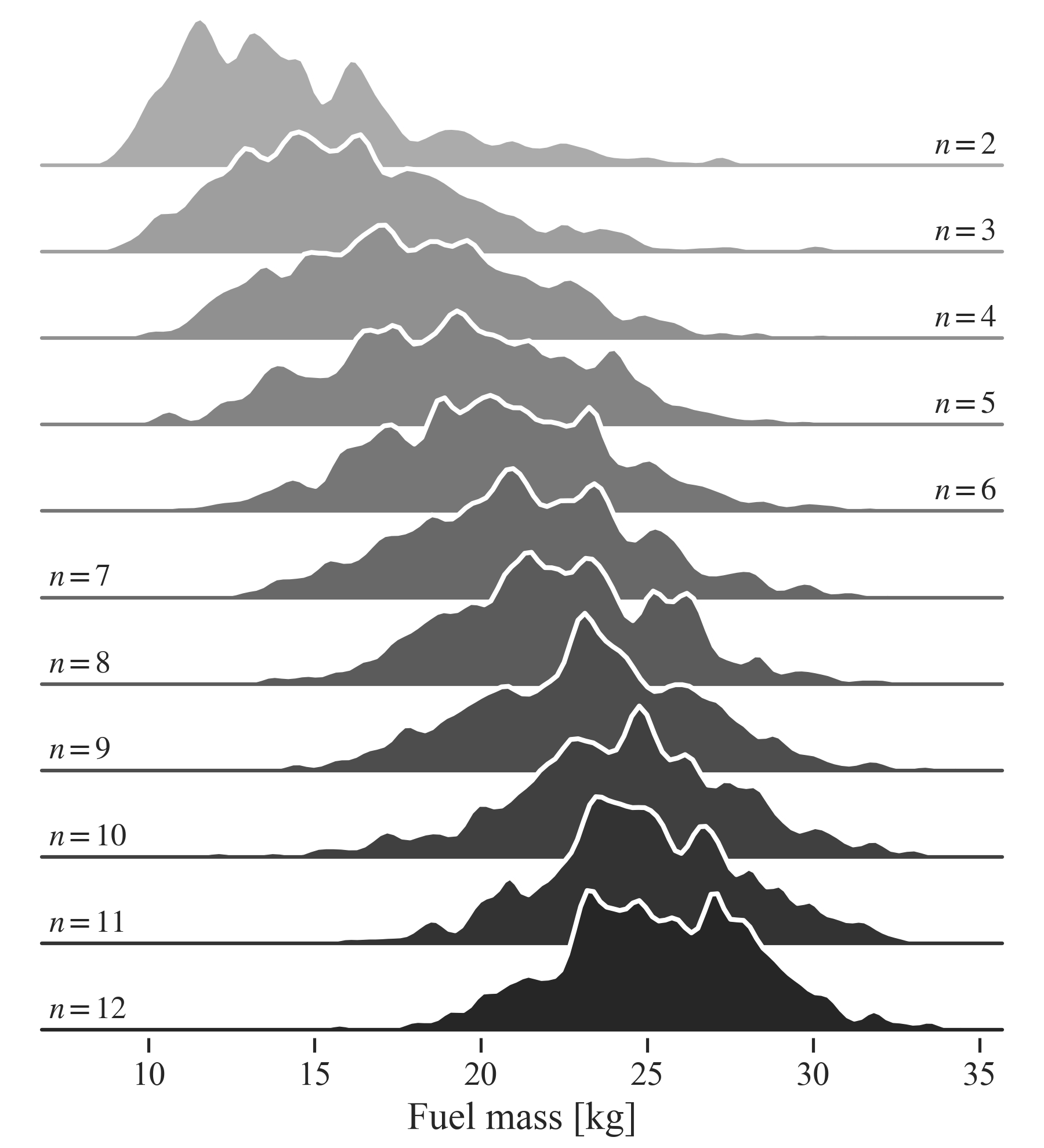}
    \caption{Fuel consumption as a function of number of bundles.}
    \label{fig:joy diviison}
\end{figure}

\autoref{fig:pairplot} shows tour cost in terms of fuel mass consumption, $\Delta V$ required and TOF as a function of number of bundles, minimum payload mass, semi-major axis standard deviation and range, and inclination standard deviation and range. As expected, number of bundles and inclination range is the greatest driver of mission cost. \autoref{fig:joy diviison} desegregates fuel consumption by number of bundles, which is clearly the greatest driver of mission cost. In all cases OSSIE is, on average, capable of fulfilling its mission and decommissioning afterwards.

\subsection{SCP}
\label{subsec: results scp}

To validate the SCP algorithm for trajectory re-optimization and adaptation to spacecraft constraints, a selection of orbit transfers of the previous section are re-optimized. Table \ref{table:Table 2} gathers the test cases and Table \ref{table:Table 3} gathers the errors computed as SCP result with respect to target and the change in $\Delta V$ with respect to combinatorial problem solution. 

The results show that the SCP manages to adapt the transfer manoeuvres to the spacecraft constraints with minimum impact over injection errors or $\Delta V$. Note that the SCP is tuned to prioritize the mission objectives and encourage that output results meet the orbit injection accuracy requirements
($\Delta a \leq 10 [km], \Delta i \leq 0.1\degree$). This might result in an increment of the overall $\Delta V$ cost if compared to the combinatorial solution, as seen in cases 1 and 3 of Table \ref{table:Table 3}, even if control cost is minimized.

For the non-coplanar scenarios, i.e., change of height followed by change of altitude, the SCP manages to reduce the propellant consumption by $8.5357\%$ compared with the combinatorial approximation. Note, however, that complete eccentricity elimination is not achieved: the latter can be solved, if required, by orbit circularization procedures, which would hinder propellant cost reductions gained by the SCP. 

\begin{figure}
   \centering
   \includegraphics[width=\linewidth]{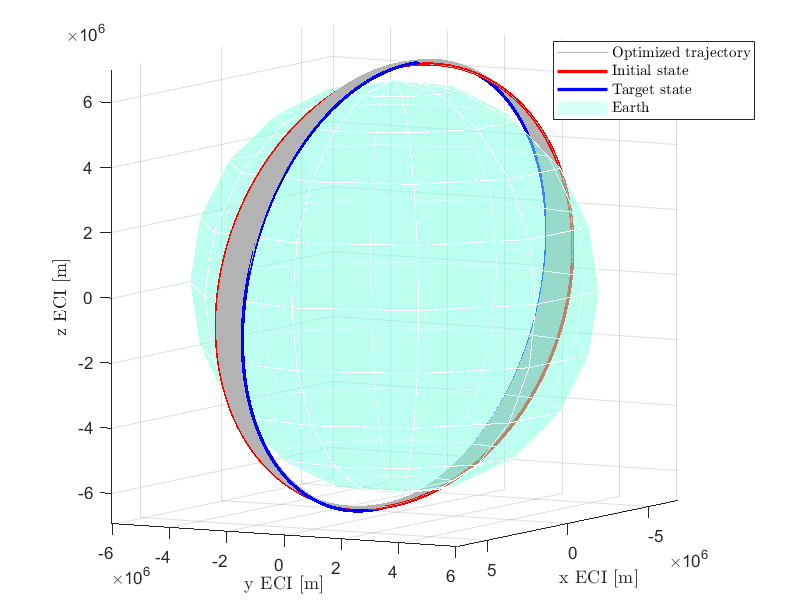}
   \caption{SCP-tailored noncoplanar maneuver (corresponding to the Test Case 2 in Table \ref{table:Table 2})}
   \label{fig:noncoplanar transfer}
\end{figure}

In general, we have observed that the current tuning benefits from a division of transfer arcs without mixing inclination and altitude change objectives simultaneously. This might be driven by the warm start that assumes such specific procedure. Other warm start strategies or SCP tunings are currently being explored for further propellant minimization by combination of objective changes. A 3D overview of the optimized transfer manoeuvre is shown in Figure \ref{fig:noncoplanar transfer}. 

\begin{table*}[t]
\caption{Summary of OSSIE scenarios under analysis: 1) coplanar transfer, 2) noncoplanar transfer, 3) inclination change (0.25\degree), 4) large inclination change (1\degree).}
\label{table:Table 2}
\centering
\begin{adjustbox}{max width=\linewidth}
\begin{tabular}{ llllll }
 \thickhline
 Test Case & TOF [hr] & $a_{t_{i}}$ [km] & $a_{t_{f}}$ [km] & $i_{t_{i}}$ [deg] & $i_{t_{f}}$ [deg]\\
 \hline
   Case 1 & 163.45   & 6950 & 7000 & 97.3964  & 97.3964\\
  Case 2 & 263.93   & 6950 & 7000 & 97.2714  & 97.5214\\
  Case 3 & 99.57   & 7000 & 7000 & 97.2714  & 97.5214\\
  Case 4 & 192.87  & 7000 & 7000 & 96.8964  & 97.8964\\
\thickhline
\end{tabular}
\end{adjustbox}
\end{table*}

\begin{table*}[t]
\caption{Comparison between the SCP manoeuvre arrival orbit and the desired target orbit.}
\label{table:Table 3}
\centering
\begin{adjustbox}{max width=\linewidth}
\begin{tabular}{ llllllll }
\thickhline
 Test Case & TOF [hr] & ${\Delta V}_\text{SCP}$ [m/s] & ${\Delta V}_\text{combin}$ [m/s] & ${\Delta V}_\text{reduct}$ [\%] & ${\Delta a}_\text{target}$ [km] & ${\Delta e}_\text{target}$ [deg] & ${\Delta i}_\text{target}$ [-]\\
 \hline
  Case 1 & 163.37  & 29.59 & 27.09 & 9.20 & 1.54 & -0.0007 & 0.0007 \\
  Case 2 & 262.43  & 29.91 & 32.71 & -8.53 & 0.06 & 0.0116 & 0.0537\\
  Case 3 & 99.47  & 32.71 & 32.71 & 0.00 & 1.57 & 0.0167 & 0.0040\\
  Case 4 & 192.87 & 132.42 & 132.72 & 0.22 & 0.05 & 0.0052 & 0.0144\\
\thickhline
\end{tabular}
\end{adjustbox}
\end{table*}

\subsection{Verification in High-fidelity Simulator}
\label{subsec: verification results}

The SENER OSSIE FES high-fidelity simulation environment serves as the core framework for modelling, testing and verifying. Matlab Simulink \cite{the_mathworks_inc_simulation_2023} is used to handle simulation, while Matlab \cite{the_mathworks_inc_matlab_2023} is used for pre- and post-processing tasks.

Translational dynamics are simulated using the Cowell propagator. Five natural perturbations are considered in the acceleration model: non-spherical Earth effects accounting for zonal harmonics of up to degree 6, atmospheric drag (density obtained from the U.S. Standard Atmosphere model), third-body perturbations from the Sun and Moon and solar radiation pressure. 
Spacecraft mass is propagated using a constant $I_{sp}$ mass propagator. \\
Attitude dynamics are modeled using Euler’s equation, which incorporates the effects of external torques: as OSSIE makes no use of internal momentum devices, internal torques are not considered in this analysis as they have a negligible magnitude in the actual spacecraft configuration. 
The key external torques considered are gravity gradient (modeled using the J2 gravitational coefficient), magnetic torques (perturbation by means of residual magnetic dipole moment), aerodynamic drag, and solar radiation pressure. Other potential torque sources, such as control system torques by means of the reaction control system, are modeled according to the precise thrust curves provided by the actuator constructed models. \\
The simulation is integrated using a fixed-step \ordinalnum{4} order Runge-Kutta integrator running at 200 Hz (timestep of 5 ms), enabling accurate system characterization while maintaining sufficient simulation speed, to perform consequent testing campaigns.

\begin{figure}
    \centering
    \includegraphics[width=\linewidth]{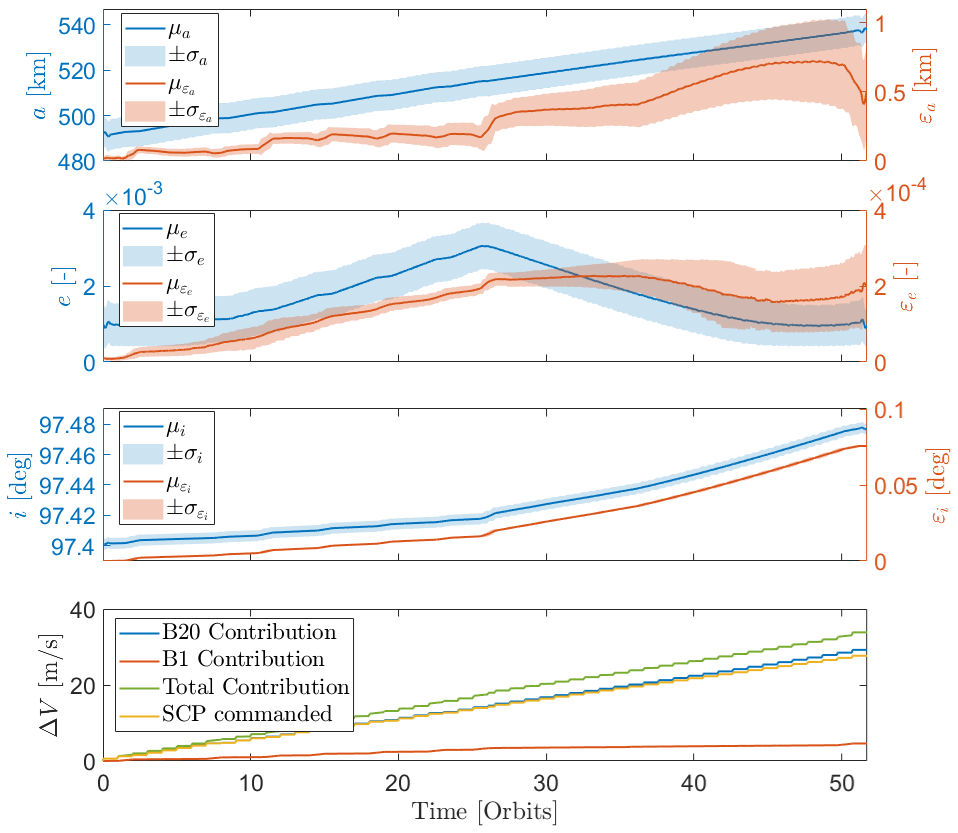}
       \caption{Preliminary verification results of case 1. Top 3 panels: moving averages, 30 minute window.}
    \label{fig:verification}
\end{figure}

The results of the preliminary verification of case 1 in \autoref{table:Table 2} in the FES are presented in \autoref{fig:verification}: the first upper three plots depict the evolution in time of the orbital parameters during the simulation (in blue), demonstrating that the optimized trajectory was correctly followed, as contrasted by the error with respect to the SCP (in orange). The test shows that the achieved error range is compliant with client requirements. The last subplot presents a comparison between the $\Delta V$ consumption achieved by the FES and the SCP: the additional $\Delta V$ requested by GNC results from pointing correction manoeuvres, to align the B20 thrusters with the thrusting direction computed in the optimization framework. 

\section{Conclusion}
\label{sec:conclusion}

We have presented a framework for addressing the Multi-target Rendezvous problem, applied to the UARX Space OSSIE mission. The framework determines optimal sequences for visiting multiple targets and generates near fuel-optimal trajectories. By integrating an Attention-based routing policy trained with Reinforcement Learning, the combinatorial optimization process is enhanced, demonstrating the effectiveness of RL in spacecraft routing. The framework accommodates the propulsion requirements of the OSSIE system, ensuring high modularity and adaptability to mission-specific constraints. This tool provides feasible and optimal guidance with minimal design effort and allows for future extensions, including the incorporation of more complex thrust profiles such as low-thrust scenarios. The proposed approach advances space vehicle routing methodologies and supports further development using reinforcement learning and advanced optimization techniques.

\section*{Acknowledgements}

This work is the result of a collaborative effort between the Delft University of Technology and SENER Aerospace \& Defence. We would like to thank Marc Naeije from TU Delft for his endorsement of the project and for his keen supervision and advice. We would like to extend our gratitude to Mercedes Ruiz for making this research possible at Sener Aerospace \& Defence. Lastly, we thank UARX Space for entrusting us with OSSIE's maiden flight, the first of many missions to come. Ad astra.

\printbibliography

\end{document}